\documentclass[10pt,twocolumn]{article}
\usepackage{epsfig}
\usepackage{harvard}

\newcommand{\be}{\begin{equation}}
\newcommand{\ee}{\end{equation}}
\newcommand{\bc}{\begin{center}}
\newcommand{\ec}{\end{center}}
\newcommand{\bi}{\begin{itemize}}
\newcommand{\ei}{\end{itemize}}
\newcommand{\ba}{\begin{eqnarray}}
\newcommand{\ea}{\end{eqnarray}}
\newcommand{\ie}{{\it i.e. }}

\newcommand{\ignore}[1]{}

\begin{document}

\title{Cultural transmission and optimization dynamics \footnote{We acknowledge financial support from MCyT (Spain) and FEDER (EU)
through Projects CONOCE and BFM2002-04474-C02-01.
}}

\author{Konstantin Klemm\footnote{Present address: Niels Bohr Institute,
Blegdamsvej 17, 2100 Copenhagen \O, Denmark}
\footnote{Corrsponding author: K. Klemm, Niels Bohr Institute,
Blegdamsvej 17, 2100 Copenhagen \O, Denmark; email: klemm@nbi.dk}
% telephone +45 353 25 273; fax 45 353 25 425; }
, V\'\i ctor M. Egu\'\i luz,
Ra\'ul Toral, Maxi San Miguel
\\Instituto Mediterr\'aneo de Estudios Avanzados IMEDEA (CSIC-UIB)
\\E-07071 Palma de Mallorca (Spain)}

\maketitle

\begin{abstract}
We study the one-dimensional version of Axelrod's model of cultural
transmission from the point of view of optimization dynamics. We show
the existence of a Lyapunov potential for the dynamics. The global
minimum of the potential, or optimum state, is the monocultural uniform
state, which is reached for an initial diversity of the population below
a critical value. Above this value, the dynamics settles in a
multicultural or polarized state. These multicultural attractors are
not local minima of the potential, so that any small perturbation
initiates the search for the optimum state. Cultural drift is modelled
by such perturbations acting at a finite rate. If the noise rate is
small, the system reaches the optimum monocultural state. However, if
the noise rate is above a critical value, that depends on the system
size, noise sustains a polarized dynamical state.
\end{abstract}

% {JEL: C63}
% \newpage
%%%%%%%%%%%%%%%%%%%%%%%%%%%%%%%%%%%%%%%%%%%%%%%%%%%%%
\section{Introduction}

Models of social dynamics are instrumental in studying mechanisms that
lead from unorganized individual actions to collective social phenomena
\cite{Schelling71,Schelling78}. In many cases the collective effects
dominate in such a way that a reductionist view in terms of individual
psychology might not be appropriate, with many detailed individual
characteristics being possibly irrelevant for the collective
macrobehavior of the system. One of the paradigms in these collective
phenomena is the study of emergence of a social consensus or uniform
state versus the emergence of a polarized state with different
coexisting social options. Examples are found among models of
segregation \cite{Schelling71}, opinion formation, dissemination of
culture \cite{Axelrod97a,Axelrod97b}, general models of social
influence \cite{Latane94} and social dynamics \cite{Epstein96}.

An important aspect in the issue of consensus versus polarization is
the spatial distribution of individuals that determines the network of
social interactions. Models that incorporate this ingredient through
local social interactions often lead to a polarized state
\cite{Latane94,Axelrod97a,Axelrod97b}: Uniformity is not reached in
spite of local mechanisms of convergence. Given the analogy with
cooperative phenomena in the Physical Sciences, ``order parameters'' have
been introduced to give quantitative measures of the order emerging in
the system \cite{Lewenstein92,Latane94}. These are global averaged
variables useful to describe changes of macrobehavior in the system.

In most studies of these phenomena the individuals are characterized by
one attribute with a two-fold option (black or white, majority or
minority viewpoint, pros and cons of some issue, etc.). An exception to
this dichotomous world is the Axelrod Culture Model (ACM) for culture
transmission \cite{Axelrod97a,Axelrod97b}. In this model culture is
defined as the set of attributes subject to social influence. Each
individual is characterized by a set of $F$ cultural {\it features},
each of which can take $q$ values that represent the possible {\it
traits} of each feature. In addition to treating culture as
multidimensional, a novelty of the model is that its dynamics takes
into account the interaction between the different cultural features.
The basic premise of the model is that the more similar an actor is to
a neighbor, the more likely the actor will adopt one of the neighbor's
traits. This similarity criterion for social influence is an example of
social comparison theory in which individuals are mostly influenced by
similar others. Processes of social influence and the origins of social
networks have been reviewed by \citeasnoun{Lazer01}. In the present paper
we will restrict ourselves to the simplest situation studied by Axelrod.
Namely, individuals are geographically distributed in the sites of a
regular grid and they interact with their immediate neighbors according
to the similarity criterion.

The ACM illustrates how local convergence can generate global
polarization. In a typical dynamical evolution the system freezes in a
multicultural state with coexisting spatial domains or clusters of
different culture. The number of these domains is taken as a measure of
cultural diversity. It is interesting to notice that it is precisely
the similarity criterion that leads to polarization and stops the
evolution towards a uniform monocultural state. Indeed, if similarity
is not taken into account to weight the probability of social
interaction, the system always reaches the consensus or uniform state
\cite{Kennedy98}. Axelrod himself explored how the number of domains in
the final state changes with the scope of cultural possibilities given
by $F$ and $q$, with the range of the interactions and with the size of
the system. The robustness of the predictions of the model has been
checked by {\it aligning} it with the Sugarscape model of
\citeasnoun{Epstein96}. In addition, the ACM has been extended in a
number of ways, which include its use as an algorithm for optimizing
cognition \cite{Kennedy98} and a study with a gradual increase of the
range of interaction \cite{Greig02}, which suggests that the increase
in communications promotes the emergence of a global but hybrid culture
rather than imposing initially dominant cultural features. The effect
of mass media in the cultural evolution has also been incorporated in
the ACM \cite{Shibanai01}, with the surprising result that mass media
promotes cultural diversity. However, this result, as some
of the others mentioned above, was obtained for a fixed set of values of
the number of features $F$ and traits $q$. It is likely that for a
different value of $q$ the effect of mass media could be the contrary.
A systematic analysis of the dependence on $q$ of the original ACM was
carried out from the point of view of Statistical Physics by
\citeasnoun{Castellano00} through extensive numerical simulations.
Defining an order parameter as the relative size of the largest
homogeneous cultural domain, these authors unveil an order-disorder
transition: There exists a threshold value $q_c$, such that for $q<q_c$
the system orders in a monocultural uniform state, while for $q>q_c$
the system freezes in a polarized or multicultural state. This result
partially modifies the original conclusions of Axelrod, in the sense
that consensus or polarization is determined by a parameter $q$ which
measures the degree of initial disorder in the system.

Cultural evolution might be thought of as an optimizing process. For
instance, it can be argued that in social comparison theory individuals
seek to optimize their social integration \cite{Lazer01}. The ACM
itself, and extensions thereof, have been shown to be able to optimize
complex functions \cite{Kennedy98}, suggesting that social interaction
is an optimization process. Likewise, the dissemination of
technological innovations can be also seen as the search for an
optimum, but in which the system can lock-in a suboptimal state
\cite{Leydesdorff01}, as in the well known example of the
QWERT-keyboard. From this perspective one might think that the
consensus or monocultural uniform state is an optimum state, while the
polarized multicultural state represent suboptimal states in which the
system gets trapped. To make such ideas on optimization dynamics
quantitative one needs to have a Lyapunov potential
\cite{Guckenheimer83} for the system dynamics. The Lyapunov potential
is a functional of the configuration of the system such that it can
only decrease or remain equal during the dynamical evolution of the
system. Although having a Lyapunov potential does not determine the
dynamics of the system \cite{Montagne96,SanMiguel00}, it can be stated
quite generally that the system will
evolve minimizing the potential until it is trapped in an attractor of
the dynamics.

In this paper we address the question of the ACM as a dynamical
optimization process by constructing a Lyapunov potential for the
system. The potential is only shown to exist for a one-dimensional
system: individuals are distributed at regular intervals along a line.
This simplest geographical set-up allows us to discuss and make clear
most of the concepts and mechanisms also occurring in higher
dimensional systems. Considering one-dimensional systems for clarity of
concepts is within the tradition of studies of models of social
dynamics \cite{Schelling71}. In addition, this one dimensional
configuration was also considered by Axelrod to exemplify {\it dialect}
dynamics. We show that the global minimum of the Lyapunov potential for
the one-dimensional ACM model is the uniform monocultural state. In
addition, we show that the potential has no other local minima. The
other attractors of the dynamics, corresponding to multicultural
states, are shown to have nearby configurations of the same or lower
value of the potential. This implies that when the system is trapped in
one of these states for $q>q_c$, any small perturbation will take the
system away from the multicultural attractor and the optimization
process will continue. Such perturbation can be seen as the effect of
cultural drift and this result answers the question posed by Axelrod
\cite{Axelrod97a,Axelrod97b}: {\it ``Perhaps the most interesting
extension and, at the same time, the most difficult to analyze is
cultural drift''}. In this sense, cultural drift, against the naive
expectation of promoting differentiation, is an efficient mechanism to
take the system to the optimum uniform monocultural state. The result
is reminiscent of the effect of randomness (named {\it temperature}) in
the studies of social impact theory \cite{Latane94} which was also
shown to increase the self-organization tendencies in the system.

The paper is organized as follows. Section 2 reviews the formal
definitions involved in the ACM model. Section 3 describes the
main features of the ACM in a one dimensional world, including the
classification of dynamical attractors and the order-disorder
transition observed for a threshold value of $q$. In section 4 we
introduce the Lyapunov potential and we use it to characterize the
order-disorder transition and the stability of the multicultural
or polarized attractors of the dynamics. Any perturbation acting
on these states is shown to take the system to the uniform
monocultural state. Section 5 is devoted to a discussion of
cultural drift. Concluding remarks and an outlook of this work is
given in section 6.

%%%%%%%%%%%%%%%%%%%%%%%%%%%%%%%%%%%%%%%%%%%%%%%%%%%%%%%%%%%%%%%%%%%%%%%
\section{Axelrod model}

The model we study is defined \cite{Axelrod97a,Axelrod97b} by
considering $N$ individuals or agents as the sites of a network. The
state of agent $i$ is a vector of $F$ components ({\em cultural
features}) $(\sigma_{i1},\sigma_{i2},\cdots,\sigma_{iF})$. Each
$\sigma_{if}$ is one of the $q$ integer values ({\em cultural traits})
$1,\dots,q$, initially assigned independently and with equal
probability $1/q$. The time-discrete dynamics is defined as iterating
the following steps:
\begin{enumerate}
\item Select at random a pair of sites of the network connected by a bond $(i,j)$.
\item Calculate the {\em overlap} (number of shared features)
$l(i,j) = \sum_{f=1}^F \delta_{\sigma_{if},\sigma_{jf}}$.
\item If $0<l(i,j)<F$, the bond is said to be {\sl active} and sites
$i$ and $j$ interact with probability $l(i,j)/F$. In case of interaction,
choose $g$ randomly such that $\sigma_{ig}\neq\sigma_{jg}$ and set $\sigma_{ig}=\sigma_{jg}$.
\end{enumerate}

In any finite network the dynamics settles into an {\em absorbing}
state, characterized by the absence of active bonds. Obviously all the
completely homogeneous configurations are absorbing. Homogeneous means
here that all the sites have the same value of the cultural trait for
each cultural feature. Inhomogeneous states consisting of two or more
homogeneous domains separated by inactive bonds with zero overlap are
absorbing as well. A domain is here defined by a set of sites connected
by bonds.

\ignore{
An order parameter in this system can be
defined \cite{Castellano00} as the relative size of the largest
homogeneous domain $S_{\rm max}/N$, being $N$ the number of sites in
the network. The control parameter for the non-equilibrium transition
described in \cite{Castellano00} is the number of traits $q$, which
measures the degree of disorder of the initial random configuration. In
a regular two dimensional network the average order parameter becomes
of order one for $q<q_c$ (ordered state) and tends to zero for $q>q_c$
(disordered state).}

%%%%%%%%%%%%%%%%%%%%%%%%%%%%%%%%%%%%%%%%%%%%%%%%%%%%%%%%%%%%%%
\section{A one-dimensional world}

We consider the case of a one-dimensional lattice formed by $N$
agents with first neighbors interaction with open boundary conditions.
Each agent $i$ can only interact with his right $i+1$ and left 
$i-1$ neighbors. We define a {\em cultural domain} as a contiguous set of agents
with the same cultural traits for all the features. Then the system
settles in an absorbing state consisting of cultural domains separated
by bonds with no overlap. These constitute the {\em barriers} through
which no interaction occurs. Thus the absorbing states can be
classified according to the number of barriers or equivalently by the
number of cultural domains. A polarized or multicultural configuration
corresponds to an absorbing state containing several cultural domains
while a uniform or monocultural state is formed by a single culture
spanning along all the sites of the lattice.

%%%%%%%%%%%%%%%%%%%%%%%%%%%%%%%%%%%%%%%%%%%%%%%%%%%%%%%%%%%Fig 1
\begin{figure}
\centerline{\epsfig{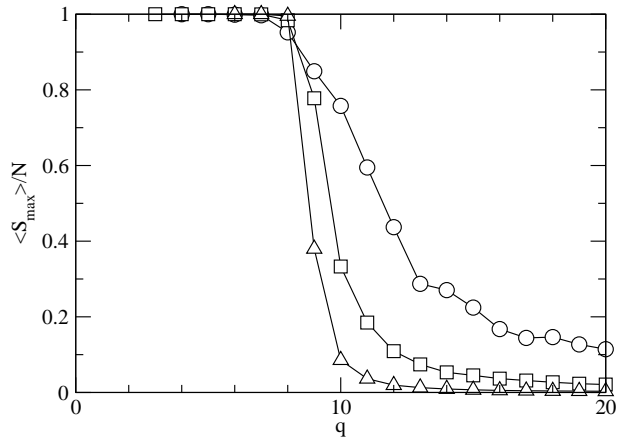}}
\caption{\label{f1d} The average order parameter $<S_{\rm max}>/
N$ in one-dimensional lattices as a function of $q$ for system
sizes $N=100$ (circles), 1000 (squares), 10000 (triangles). Each
plotted value is an average over 100 runs with independent initial
conditions. Number of features $F=10$. }
\end{figure}

Extensive numerical simulation show that in not too small systems only
monocultural or extremely polarized configurations are reached. This
behavior is captured by an order parameter defined here as the relative
size of the largest homogeneous cultural domain $S_{\rm max} / N$.
Clearly, if this quantity is unity, one culture spans the whole system,
corresponding to a monocultural state. On the other hand, if none of
the cultural domains reaches a size that is visible on the scale of the
system size, $S_{\rm max}\ll N$, the configuration is extremely
polarized. Then an agent shares his cultural attributes with only a
small neighborhood.

\begin{figure}
\centerline{\epsfig{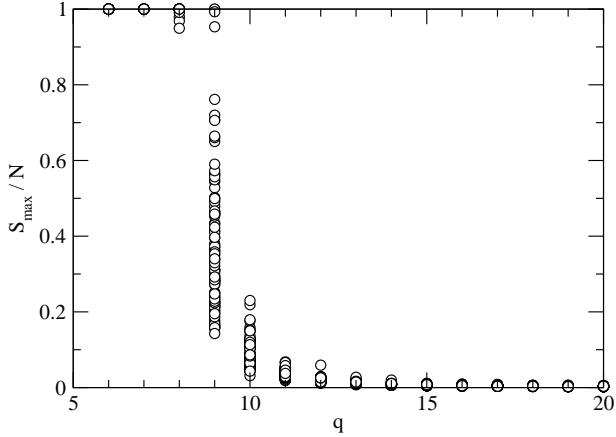}}
\caption{\label{f1dr} Scatter plot of the order parameter in one-dimensional
lattices as a function of $q$ for system size $N=1000$ and $F=10$ features.
For each value of $q$ the outcome of 100 independent runs is plotted.
}
\end{figure}

Figure\ \ref{f1d} shows, for $F=10$, the values of the average order
parameter $\langle S_{\rm max}\rangle / N$ in the final absorbing state
as a function of the number of available traits $q$. For $q<8$ we
always find a monocultural absorbing state. Increasing $q$ beyond 8,
$\langle S_{\rm max} \rangle / N$ drops towards zero, the more rapidly
the larger the system, indicating the existence of a transition for $q
\simeq 8$. This change of behavior between monocultural and polarized
states is emphasized when looking at the outcomes of the realizations
themselves (without averaging) in Fig.~\ref{f1dr}. We observe that this
transition is not accompanied by a regime of bistability close to
$q_c$. This means that it does not exist a finite range of $q$-values
for which a similar number of realizations finish either in the
monocultural or in the multicultural state. The absence of bistability
suggests that the transition can be classified as continuous,
while a similar type of transition observed in two-dimensional lattices
is accompanied by a bistable regime \cite{Castellano00,Klemm02b},
indicating that in general the transition is discontinuous or first
order.

It is important here to note that the control parameter $q$ that
governs this  transition or change of behavior is not a parameter that
can be tuned in a given system. Rather it enters in the definition of
the system, and therefore the transition corresponds to a change of
behavior in a class of systems that we explore by changing $q$.  On the
other hand, the dynamic rules do not change with $q$ and the crucial
way through which $q$ enters in the dynamic evolution is in the initial
condition. We have chosen random initial conditions with a uniform
probability distribution for the value taken by each feature. With this
choice, $q$ gives a measure of initial disorder. The transition that we
discuss refers to an average behavior, the average being taken over an
ensemble of such initial conditions, and the value of $q_c$ reflects
this choice of initial conditions. For initial conditions with large
initial disorder $q>q_c$ the system freezes in a multicultural
configuration, while for a small initial disorder ($q < q_c$) the
system reaches the monocultural state. If a different choice of random
initial conditions is made, for example taking a Poisson distribution
\cite{Castellano00}, the value of $q_c$ changes. Of course, if the
ensemble of initial conditions were restricted, for instance, to
homogeneous states, no transition would be observed.

%%%%%%%%%%%%%%%%%%%%%%%%%%%%%%%%%%%%%%%%%%%%%%%%%%%%%%%%Fig 2

%%%%%%%%%%%%%%%%%%%%%%%%%%%%%%%%%%%%%%%%%%%%%%%%%%%%%%%%%%%%%%%%%%%%%
\section{Lyapunov potential}

In the one-dimensional version of the model the dynamics can be
described in terms of a {\em Lyapunov potential}. A function of the
state of the system $L(\{\sigma\})$ is a Lyapunov potential if its
value does not increase during the dynamical evolution. If $L(t)$
represents the Lyapunov potential of the system at time $t$, then $L(t)
\ge L(t+1)$. We state that the negative {\em total overlap}
\be
L = - \sum_{i=1}^{N} l(i,i+1)~,
\ee
is a Lyapunov potential of the one-dimensional
Axelrod model.

\begin{figure}
\centerline{\epsfig{file=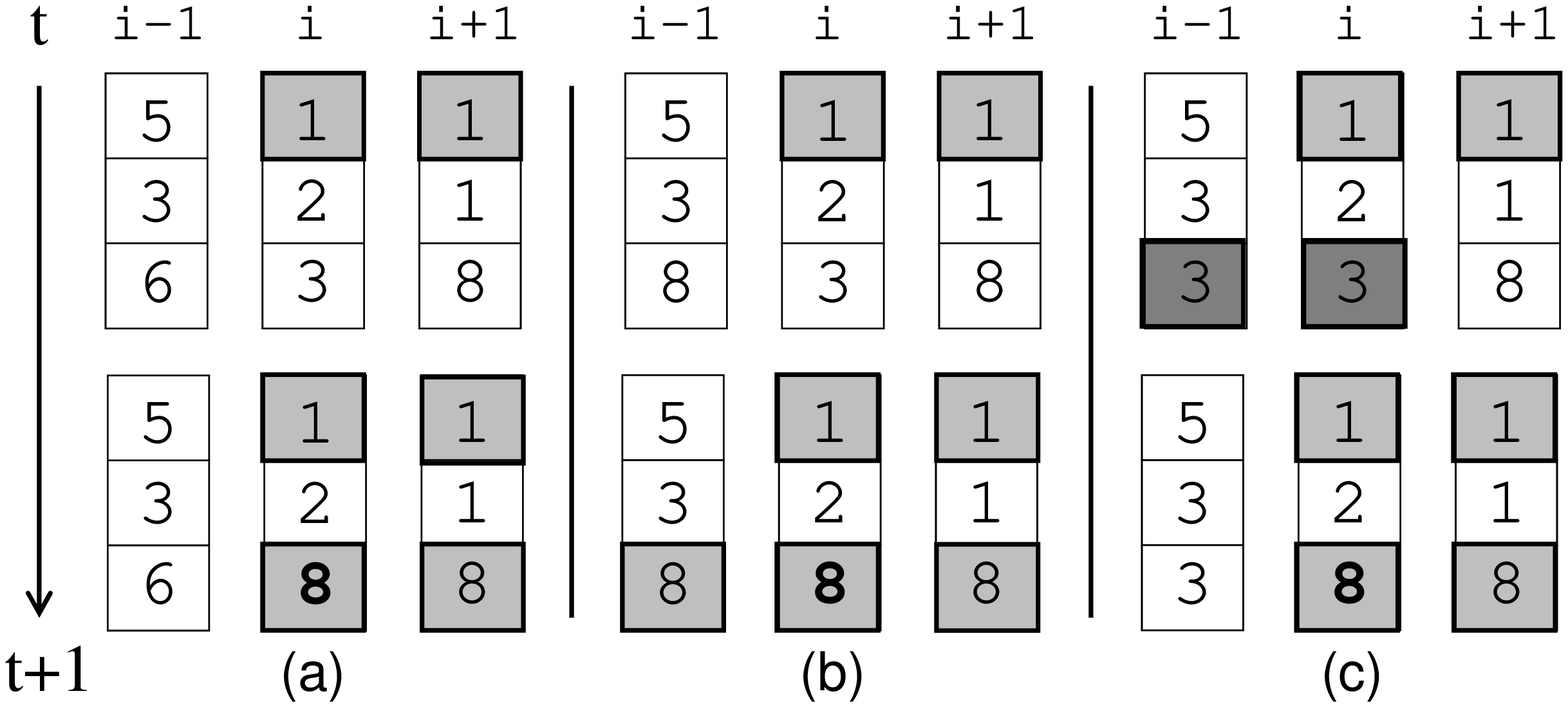,width=.45\textwidth,angle=-0}}
\vskip -2cm
\caption{\label{flya} Three possible outcomes of an interaction between
agents $i$ and $i+1$ for a system with $F=3$ features and $q=10$.
Shared features are indicated by grey background. The trait of feature
$\sigma_{i3} = 3$ is switched to $\sigma_{i3} = \sigma_{(i+1)3} =
8$. The new acquired trait by agent $i$ increases the overlap with its
$i+1$ neighbor but (a) has no effect on $i-1$ [$L(t+1) = L(t)-1$]; (b)
increases the overlap with $i-1$ [$L(t+1) = L(t)-2$]; (c) decreases the
overlap with $i-1$ [$L(t+1) = L(t)$]. }
\end{figure}

{\bf Proof}: We have to show that the negative total overlap cannot
be increases by an interaction\footnote{If there is no interaction, the
state of the system does not change and then the Lyapunov potential
remains unchanged}. At a given time step the bond $(i,i+1)$ is selected
so that agent $i$ acquires one of the traits of agent $i+1$ \footnote{In
the example in Fig.~\ref{flya}, $F=3$ and $\sigma_{i2} (t+1)=
\sigma_{(i+1)2} (t) = 8$} Then the overlap across that bond increases
by one unit. For the overlap across the other bond $(i-1,i)$ of site $i$
there are three possibilities (see Fig.~\ref{flya}):
\begin{enumerate}
\item it is the same as before, if the acquired trait is shared by
agent $i-1$ or the discarded trait was shared by agent $i-1$. Then
$L(t+1)=L(t) -1$ (Fig.~\ref{flya}a), \item it increases by one unit,
if the acquired trait is also shared by the agent $i-1$. Then
$L(t+1)=L(t) -2$ (Fig.~\ref{flya}b), or
\item it decreases by one unit, if the change occurred with respect to
one of the shared traits with $i-1$. Then $L(t+1)=L(t)$
(Fig.~\ref{flya}c).
\end{enumerate}
Thus an interaction the value of $L$
will be less than before or the same as before. Taking into account
that all other terms in $L$ do not vary, we find that in any
interaction, $L$ never increases. {\bf End of proof.}

%% Lya%%%%%%%%%%%%%%%%%%%%%%%%%%%%%%%%%%%%%%%%%%%%%%%%%%%%%Fig 3

It is also convenient to relate the Lyapunov potential to the number
of bonds $n_k$ with overlap $k$:
\be
L = - \sum_{k=0}^{k=F} n_k k~,
\ee
with $\sum_{k=0}^{F} n_k = N$.
The absorbing configurations correspond to the case $n_k=0$, for
$0<k<F$. For these configurations we obtain
\be
L_{\rm absorbing} = - n_F F = -(N - n_0) F~,
\ee
where $n_0$ is the number of bonds with zero overlap, that is the
number of barriers. Therefore, the absorbing states can be ordered
according to the number of barriers. In the monocultural homogeneous
states all the bonds have overlap $F$ and thus $n_k = 0$, $\forall k
\ne F$ and $n_F = N$. They correspond to the absolute minima of the
Lyapunov potential or optimum states with $L_0= - N F$. There is a
multiplicity of these minima corresponding to the $\eta_0=q^F$
equivalent different cultures ($\sigma_{if}=\sigma_{jf}$ $\forall i,j$
and $f$), characterized by the combination of cultural traits. It is
important to notice that they are equivalent. Which of the optimum
states is selected for $q<q_c$ depends on the initial conditions and
the stochastic realization of the dynamics.

Inhomogeneous multicultural states consisting of two or more
homogeneous domains separated by barriers are absorbing as well. We can
order the multicultural absorbing configurations according to their
Lyapunov potential. The first absorbing configurations (different from
the monocultural states) in potential correspond to the coexistence of
two different cultural domains separated by one barrier. There are
$\eta_1=[q(q-1)]^FN$ configurations with a value of the Lyapunov
potential $L_1 = -(N-1)F$. In this case $n_0 = 1$, $n_F=N-1$ and all
other $n_k = 0$. The next level corresponds to three cultural domains
(and two bonds of zero overlap), with a potential $L_2= - (N-2) F$
($n_0 = 2$, $n_F=N-2$ and all other $n_k = 0$ and $\eta_2 =
[q(q-1)^2]^F N (N-1)/2$ equivalent configurations. In general an
absorbing state with $K+1$ cultural domains and $K$ barriers will have
\be
L_K = - (N-K) F~,
\ee and
\be
\eta_K = [q(q-1)^K]^F \left( \begin{array}{c}
  N \\
  K
\end{array}\right)
\ee equivalent configurations.

%%%%%%%%%%%%%%%%%%%%%%%%%%%%%%%%%%%%%%%%%%%%%%%%%%%%%%%%%%%%%Fig 4
\begin{figure}
\centerline{\epsfig{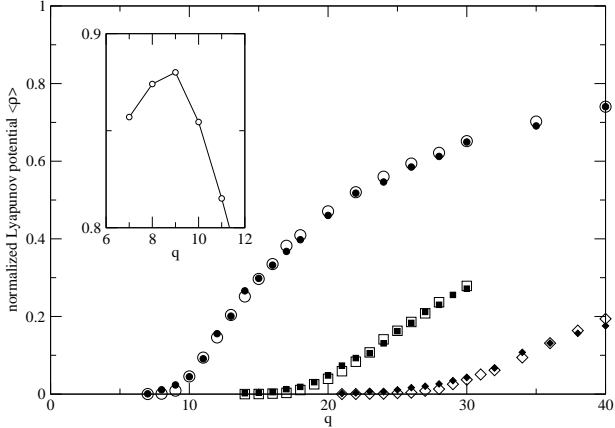}}
\caption{\label{fdens} Normalized Lyapunov potential in the absorbing state
as a function of $q$ for $F=10$ (circles),
$F=20$ (squares) and $F=30$ (diamonds). System sizes are $N=100$
(filled symbols) and $N=1000$ (open symbols). The inset shows the
difference in the normalized Lyapunov potential between the initial value
and the value reached in the absorbing state for $N=1000$ and $F=10$.
}
\end{figure}

\begin{figure}
\centerline{\epsfig{file=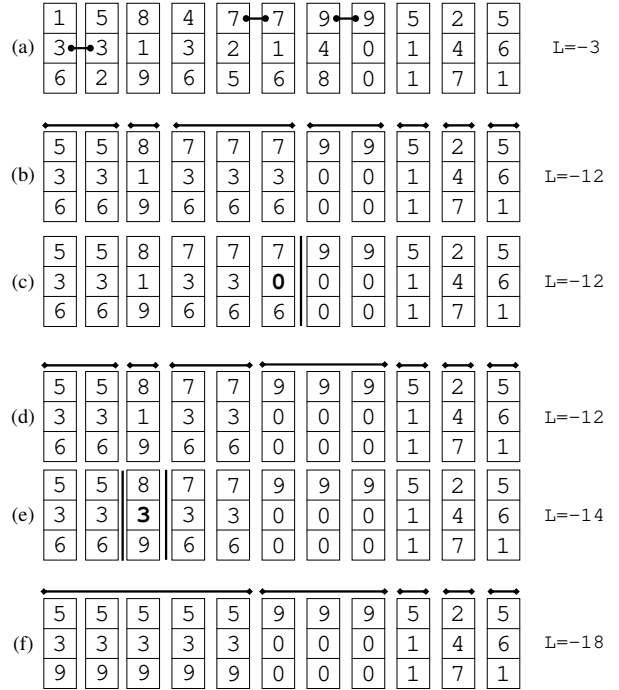,width=.45\textwidth,angle=-0}}
\vskip -1cm
\caption{\label{flya_sfp} (a) Initial condition for a system formed by
$N=10$ agents, with $F=3$ features, and $q=10$. The average potential
for the initial condition is $\langle L \rangle_0 = - NF/q$. For this
particular realization L = - 3 in accordance with the average
estimation. The overlap is indicated by the horizontal lines.
(b) The dynamics leads to an attractor which is a multicultural
configuration ($q > q_c = 3$) formed by 7 cultural domains indicated by
the bar on the top. For this configuration $L= - 12$.
(c) An exogenous perturbation switches on of the traits (indicated in
bold). The Lyapunov does not change but it opens the possibility to
interact with its right neighbor.
(d) After the evolution the new absorbing configuration is an equivalent
configuration (it also contains 7 cultural domains) with the same
value of the Lyapunov potential.
(e) A new perturbation switches one trait that decreases the potential.
(f) The system cannot come back to a state with higher potential but
reaches a new configuration of 5 cultural domains with $L = -18$.}
\end{figure}

The properties of the Lyapunov potential can give valuable insight into
the dynamics of the system. For instance, the minima of the Lyapunov
potential are absorbing states of the dynamics. However the opposite is
not true. As the dynamics never increases the potential, once a minimum
is reached the dynamics stops there because any neighboring
configuration has a larger potential. However, an absorbing
configuration can have neighboring configurations with lower or equal
potential (see Fig.~\ref{flya_sfp}, described below). The reason why
the dynamics stops in such absorbing states is not included in the
potential but in the dynamical rules.

Having characterized the absorbing configurations in terms of the
Lyapunov potential now we turn our attention to the description of the
transition in terms of the potential. In order to facilitate systematic
comparison between systems with different sizes $N$ and different numbers
of features $F$, we use the normalized Lyapunov potential 
$\rho = (L -  L_0) / NF = L / NF +1$ with values ranging from zero to
unity. In absorbing configurations $\rho$ is simply the density of barriers.
For instance, $\rho=6/100$ for a configuration with 6 barriers in a system of
size $N=100$.

\begin{figure}
\centerline{\epsfig{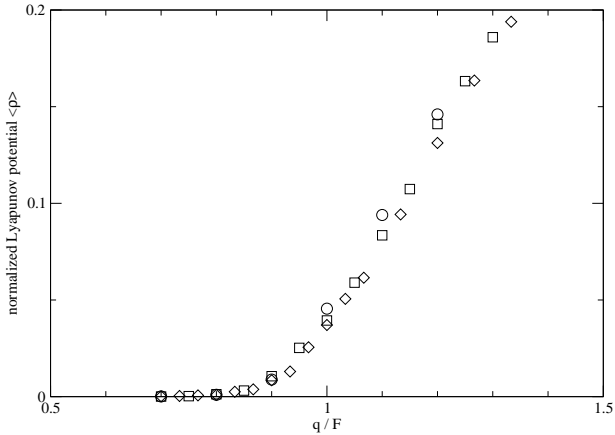}}
\caption{\label{fdens_sca} The data from Fig.~\ref{fdens} with $N=1000$ and
$F=10,20,30$ collapse when plotted as a function of the rescaled
parameter $q/F$.}
\end{figure}

\begin{figure}
\centerline{\epsfig{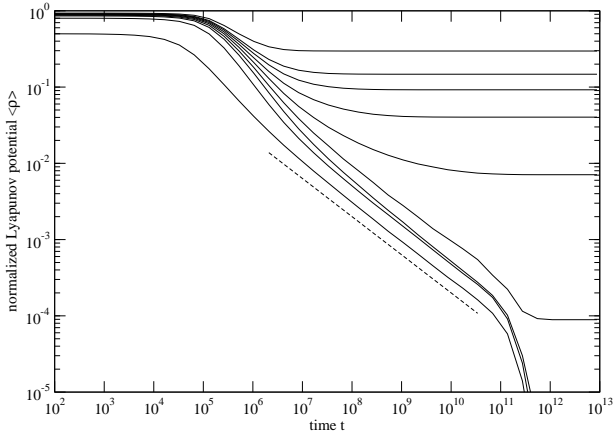}}
\caption{\label{fevol} Time evolution of the normalized Lyapunov potential for
$F=10$ and $q=2,5,7,8,9,10,11,12,15$ (solid curves, bottom to top) in
systems of size $N=10000$. For $q<q_c \approx 8$ the normalized Lyapunov
potential approaches zero according to a power law. The dashed line
has slope $-0.5$.}
\end{figure}

In Fig.~\ref{fdens} we show the average value of $\rho$ in the absorbing
state reached by the dynamics. The average is take over 100 simulations
for each data point. We observe
that, as a function of $q$, the Lyapunov potential increases
continuously from the value of a monocultural configuration,
incorporating barriers, and thus increasing the potential and $\rho$ as
$q$ increases. It is apparent in this figure that a change of behavior
between the monocultural state with $\langle L \rangle = -NF$, $\rho=0$
and the multicultural states occurs for $F\simeq q$, in agreement with
the result in Fig.~\ref{f1d}. This change of behavior also manifests
itself in the difference between the value of $\rho$ in the initial
random configuration and the final absorbing state. For the average
over our set of random initial conditions $\rho =1-1/q$. The
difference with the final value shows a maximum for $F\simeq q$ (see
inset of Fig.~\ref{fdens}). We have checked that the value of the
maximum increases linearly with $F$. The fact  that $F$ and $q$ are not
two independent relevant parameters, but that rather the scaling
parameter $F/q$ is the proper one to describe these phenomena is made
more explicit in Fig.\ \ref{fdens_sca}. In this figure the x-axis has
been rescaled as $q/F$ for the same data as in Fig.\ \ref{fdens}. We
observe a scaling phenomena in the sense that $\rho$ is seen to be a
function only of $q/F$. These results, together with Fig.\ \ref{f1d},
suggests the existence of a transition for $q=q_c\simeq F$. To make
this statement rigorous we have to find some singular behavior. We find
such behavior in the dynamical evolution of the normalized Lyapunov
potential $\rho (t)$ (Fig.\ \ref{fevol}).
In the early steps of the dynamical evolution $\rho(t)$ remains
constant or decays slowly for a large number of iterations. For values
of $q$ below the transition point $q_c$ the density decays as $\rho (t)
\sim t^{-0.5}$. For $q$ above the transition point $q_c$, $\rho$
saturates at a finite value. For $q>q_c$ an absorbing state is reached
after a time span several orders of magnitude shorter than in the case
$q<q_c$. This clearly identifies the continous transition from a
monocultural to a multicultural state for a critical value of
$q=q_c\simeq F$.

%% 1d %%%%%%%%%%%%%%%%%%%%%%%%%%%%%%%%%%%%%%%%%%%%%%%%%%%%%%%%%%Fig 5

Is the Lyapunov potential useful in understanding why the system is
trapped in a multicultural configuration for $q>q_c$? Due to the
multiplicity of configurations with a given a value of the potential
one might conjecture that the system is trapped in an absorbing
configuration which is close to the initial configuration, making a
``short excursion''. Consequently, the difference of the Lyapunov
potential in the random initial and the absorbing configuration should
be small. However, we observe that it assumes large values and has a
peak at exactly the transition point $q_c$ (see inset in Fig.\
\ref{fdens}). This fact rules out the argument that the excursion is
short in potential. In particular it is not true that when the system
reaches the optimum state is because the initial condition has a value
of the Lyapunov potential close to the optimum state. In fact for $q
\le q_c$, when the system reaches the optimum state, the excursion in
value of the potential is much larger than for $q \gg q_c$. What this
means is that  the dynamics is not gradient
\cite{Montagne96,SanMiguel00}. It does not follow the trajectory of the
steepest descent of the potential. This represents a genuine
non-equilibrium dynamics which cannot be completely described just by
the optimization of a potential. In addition, in the case considered in
this paper there are entropic contributions (the degeneracy of
equivalent configurations with the same potential) which are also at
play. Even though the dynamics is not fully determined by the
potential, the potential is very useful in understanding the stability
properties of the absorbing states.

The homogeneous configurations are local and at the same time global
minima of $L$. They are the optimum states and deviating from these in
any local step in configuration space, \ie assigning a different trait
to one of the features of an agent, always increases $L$. All other
absorbing configurations are neither local nor global minima: there are
always neighboring configurations with the same or a lower value of
$L$. From these configurations, one can always make a local step to a
non-absorbing state. From the so reached non-absorbing state the
dynamics does not necessarily return to the original absorbing state.
The dynamics does not drive spontaneously the system towards an
adjacent disordered absorbing configuration when the system is trapped
in an absorbing state. These excursions are caused by exogenous
perturbations that randomly flip a cultural trait. In terms of the
classification of stationary states the absorbing monocultural states
are stable. Properly, they are  unstable since a perturbation can take
them to a state of higher potential and from there to an equivalent
monocultural state which is also optimum. The absorbing multicultural
states are not meta-stable. There exist perturbations of the smallest
size  (a change in a single trait of an agent) that take the system to
a configuration of the same or lower value of the potential. In the
first case we talk about marginally stable states and in the second of
unstable states. Therefore the disordered absorbing configurations are
meta-stable.

We are going to illustrate these properties with the help of Fig.\
\ref{flya_sfp}. For the sake of concreteness we consider 11 agents
interacting through $N=10$ bonds with $F=3$ and $q=10$. For this case
we know that $q > q_c \simeq 3$. Thus the system should reach a
disordered configuration. The overlap between the state of the agents
in the initial condition is indicated in the figure giving $L=-3$ in
accordance with the average estimation $\langle L \rangle = - NF / q =
3$.
The second row (b) shows the final configuration after the evolution of
the system. The multicultural configuration is formed by 7 cultural
domains (indicated in the figure) having 6 barriers. Thus the Lyapunov
potential of this absorbing configuration is $L = -(10- 6) 3 = -12$.
Now an exogenous perturbation switches $\sigma_{62}$ (indicated in bold
in the third row) activating the bond with its right neighbor. Note
that this perturbation has not changed the Lyapunov potential.
Thus it is a neighboring configuration (c) with the same Lyapunov
potential: if we now let the system relax, it reaches an equivalent
configuration (7 domains) with the same potential (d).
Thus an exogenous perturbation has
lead the system to another equivalent configuration without a
modification of the potential. However, there are other perturbations
that can decrease the potential as indicated in the next row. The new
perturbation ($\sigma_{32}$) activates two bonds and then $L = -14$ (e).
Therefore the absorbing state (d) is unstable.
As
the potential cannot increase the new evolution cannot recover the
previous potential level. Instead an absorbing configuration with lower
potential is reached (f).
This configuration reached is composed of 5
domains and has Lyapunov potential $L=-18$. One may expect that by repetition of these cycles of
perturbation-relaxation the number of domains and the Lyapunov
potential are reduced further until a homogeneous configuration is
reached.

%% 1d %%%%%%%%%%%%%%%%%%%%%%%%%%%%%%%%%%%%%%%%%%%%%%%%%%%%%%Fig 6

\begin{figure}
\centerline{\epsfig{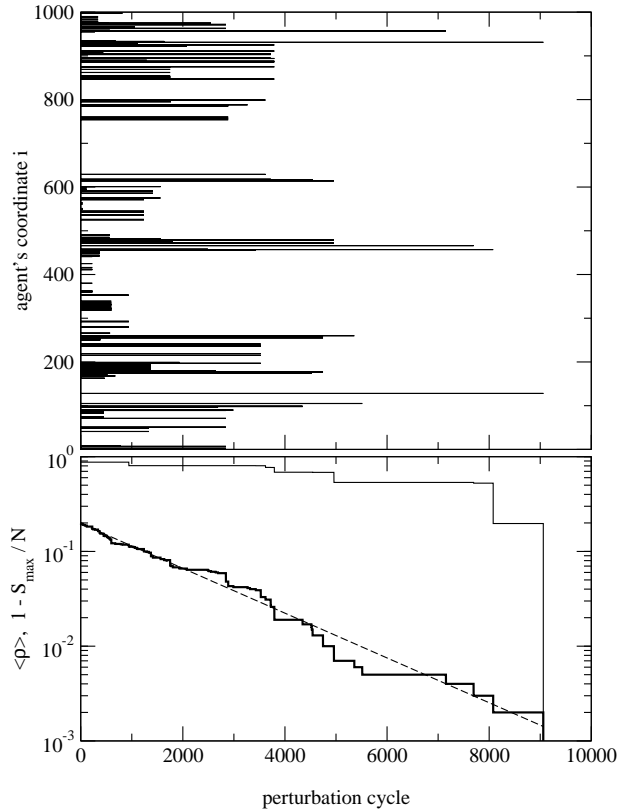}}
\caption{\label{fmetas} Ordering of the system by iterated cycles of
perturbation and relaxation. The upper panel shows the cultural
barriers (bonds with zero overlap) after a given perturbation cycle.
For the same dynamical run the lower panel shows the values of the
order parameter $S_{\rm max}$ (thin curve) and the normalized Lyapunov
potential $L$ (thick curve). Parameter choices are $F=10$, $q=13$ and
$N=1000$. }
\end{figure}

In order to analyze the consequences of the lack of stability of the
absorbing multicultural states, we have devised simulations of the
model including exogenous perturbations. The absorbing states are
subject to {\em single feature perturbations}, defined as randomly
choosing $i\in\{1,\dots,N\}$, $f\in\{1,\dots,F\}$ and
$s\in\{1,\dots,q\}$ and setting $\sigma_{if}=s$. Then the simulations
are designed as follows:
\begin{enumerate}
\item[(A)] Draw a random initial configuration.
\item[(B)] Run the dynamics by iterating
steps (1), (2) and (3), until an absorbing state is reached.
\item[(C)] Perform a single feature perturbation of the absorbing state and
resume at (B).
\end{enumerate}
In other words, whenever an absorbing configuration has been
reached, we measure $L$ and $S_{\rm max}$, perform a perturbation
and restart the dynamics from the perturbed configuration. This mimics
the effect of a random influence on the system which acts much more seldomly
than the dynamics of cultural imitation in the original model. We find
that in this case the system is driven to complete order,
i.e. $L$ gradually decreases to the minimum value $-NF$ and
$S_{\rm max}$ gradually increases to the maximum value $N$. For a
typical simulation run, Fig.\ \ref{fmetas} displays the
evolution. One observes that the normalized Lyapunov potential decays
exponentially. Recalling that in absorbing states the normalized Lyapunov
potential is simply the fraction of bonds that constitute a barrier, we see
that the number of barriers decreases exponentially.
The probability for
a given barrier to vanish during a perturbation cycle is constant, i.e.\ it
does not depend on the number of barriers present in the system.
Barriers dissolve independently.

%% Lya%%%%%%%%%%%%%%%%%%%%%%%%%%%%%%%%%%%%%%%%%%%%%%%%%%%%%%%Fig 7

From the explicit time evolution of all the barriers in the system
(upper panel of Fig.\ \ref{fmetas}) it is also apparent that no new
barriers are created. This can be understood geometrically: Let us
first consider only one arbitrary feature $f$. An interface within the
feature $f$ is a bond $(i,i+1)$ with disagreeing traits $\sigma_{i,f}
\neq \sigma_{i+1,f}$. The dynamics does not create new interfaces. When
agent $i$ adopts the trait of agent $i+1$, the interface merely moves
from $(i,i+1)$ to $(i-1,i)$. If the latter bond has been an interface
already, the two interfaces either merge or annihilate. Considering the
whole system, in an absorbing state the interfaces are the same for all
features. Trivially, all features have exactly the same number of
interfaces. In order to increase the number of barriers by perturbation
and subsequent relaxation the number of interfaces would equally have
to increase in all features. However, all but one feature (the one in
which the perturbation is performed) are guaranteed not to increase the
number of interfaces, as shown before. So the number of barriers and
the number of cultural domains do not increase. This again proves that
only configurations without barriers are stable. Once such a homogeneous
configuration has been reached, the perturbations cannot drive the
system to a different absorbing configuration. In consequence, all but
the completely ordered absorbing configurations are not stable, meaning
that minimal perturbations drive the system away from these states.

We mention again that the system is always allowed to relax to an
absorbing configuration before a perturbation is performed. However,
when perturbations occur simultaneously with the original dynamics,
their effects may accumulate. At a sufficiently large rate of
perturbations this may result in a disordered system with many
cultures. On the other hand, for very low rate of perturbations, the
scenario will be close to the alternating perturbation and relaxation
studied in this section above, resulting in a homogeneous system. The
following section is dedicated to the study of this case of ongoing
perturbations at different rates.

%%%%%%%%%%%%%%%%%%%%%%%%%%%%%%%%%%%%%%%%%%%%%%%%%%%%%%%%%%%%%%%%%%%%%
\section{Cultural drift}

In this section we address the role that cultural drift has on the
behavior of Axelrod's model. The previous section has shown that
even infinitesimal noise has a non--trivial effect and, therefore,
we expect that cultural drift, modelled as random perturbations
acting at a constant rate $r$, will have a relevant role in the
model. To be more specific, we implement cultural drift by adding
a fourth step in the iterated loop of the model defined in Section\ 2:
\begin{enumerate}
\setcounter{enumi}{3}
\item With probability $r$, perform a single feature perturbation.
\end{enumerate}
This is intended to be a more realistic effect of uncertainty in
the agent's behavior. As this kind of noise acts continuously on
the dynamics, the difference with the scenario discussed in
previous section is that the system is not necessarily in an
absorbing configuration when a perturbation occurs. Therefore, it
is not straightforward to generalize the previous results based on
the existence of fixed points of the dynamics and their stability
properties, since the dynamics is not allowed to relax
to them before a perturbation acts. We will see, however, that a
simple argument based on the random walk, is able to give us some
quantitative predictions.

%% 1d %%%%%%%%%%%%%%%%%%%%%%%%%%%%%%%%%%%%%%%%%%%%%%%%%Fig 8

\begin{figure}
\centerline{\epsfig{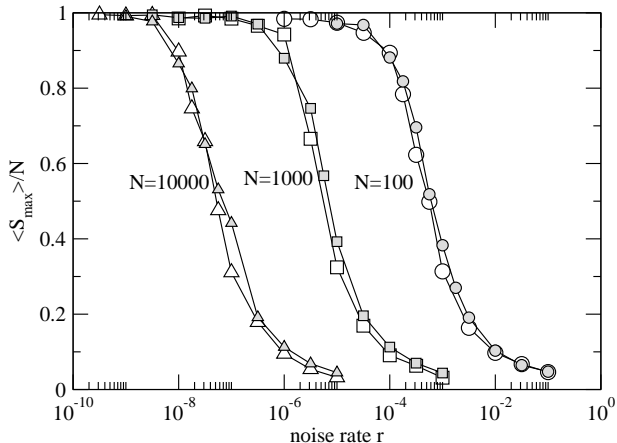}}
\caption{\label{f1dn} Dependence of the relative size of the largest cultural domain
with noise rate $r$ in one-dimensional lattices of size $N=100$
(circles), $N=1000$ (squares), $N=10000$ (diamonds), for $q=5$
(filled symbols) and $q=50$ (open symbols). Agents have $F=10$
features.}
\end{figure}

We first show the results of the numerical simulations of the
model modified to take into account the cultural drift. Figure
\ref{f1dn} shows the variation of the order parameter $\langle
S_{max}\rangle/N$ with the noise rate $r$ for different system
sizes. As expected, disorder appears for sufficiently large noise
rate $r$. The exact location of the transition point strongly
depends on the system size $N$, but it is only weakly dependent on
the number of traits $q$. In fact, the variation of $q$ from $q=5$
to $q=50$, which in the absence of noise or perturbations lead to
qualitatively different outcomes (remember that $q_c\approx 10$ in
that case), causes an almost negligible shift of the transition
towards slightly lower values of the noise rate.

%% 1d %%%%%%%%%%%%%%%%%%%%%%%%%%%%%%%%%%%%%%%%%%%%%%%%Fig 9

Although at a sufficiently low rate $r$ the situation might appear to
be close to the case of alternating perturbation and relaxation studied
in previous sections, we must stress that there is an essential
difference: for noise acting continuously the system can explore
continuously nearly homogeneous regions (with a high value of the order
parameter $\langle S_{max}\rangle/N$) by jumping from one region to
another at a time scale that grows with $N$. This reflects the
meta-stability of the different equivalent optimum states (homogeneous
cultures). We can give an intuitive explanation of the existence of a
transition from ordered to disordered states in the presence of
cultural drift: if the noise rate is such that the typical time $1/r$
between perturbations is shorter than the average relaxation time $T$,
the effect of the perturbations adds up in the system and disorder
appears. The system is in a polarized noisy dynamical state. On the
contrary, if the noise rate is small, it becomes efficient in taking
the system to explore nearby configurations of lower potential and the
optimization dynamics proceeds escaping from absorbing states. The
minima of the potential is then reached and a monocultural state
emerges. This simple picture tells us that disorder will set in when
the average relaxation time $T$ of perturbations of a homogeneous state
satisfies $rT = {\cal O} (1)$.

%% 1d %%%%%%%%%%%%%%%%%%%%%%%%%%%%%%%%%%%%%%%%%%%%%%%%%%%Fig 10
\begin{figure}
\centerline{\epsfig{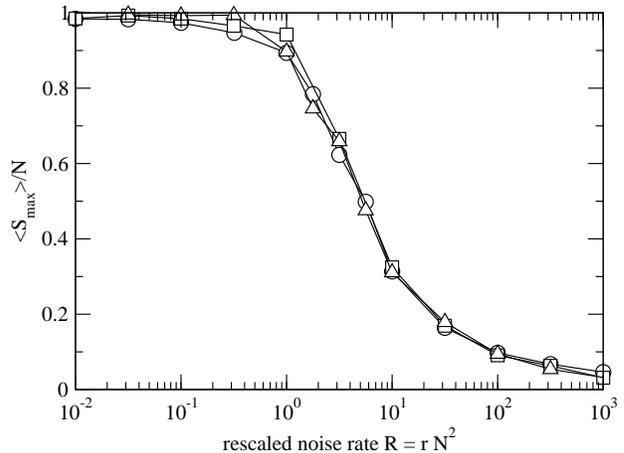}}
\caption{\label{f1ds} Scaling of the relative size of the largest
cultural domain in one-dimensional lattices. Symbols as in
Fig.\ref{f1d}, $q=50$.}
\end{figure}

It is possible to introduce an approximate argument for the
calculation of $T$. Imagine a completely ordered state as the
initial condition at $t=0$. A single feature perturbation of this
state induces a ``damage'' of size $x(t=0)=1$ in one of the
features.  In the following time steps the damage may spread until
an ordered state is reached again by $x(t)=0$ or $x(t)=N$.
Therefore we can envisage the system as a damage cluster and an
undamaged background separated by 2 active bonds
(interfaces)\footnote{We are assuming here that the system has
periodic boundary conditions. The role of the boundaries should be
negligible for sufficiently large system size $N$.}. These
interfaces execute a random walk type of diffusion and the average
time needed for them to merge in such a way that an ordered region
spans the whole system is well know \cite{Grimmett82} to scale as
\begin{equation}
T \sim N^2,
\end{equation}
so that the average relaxation time of perturbations diverges
quadratically with increasing system size.

This result is confirmed, see Fig.~\ref{f1ds}, by showing that the
data of Fig.~\ref{f1dn} collapse into a single curve when plotted
as a function of a rescaled noise rate $rN^2$, which incorporates
noise rate $r$ and system size $N$.  This shows indeed that for
increasingly larger system sizes, a vanishingly small noise rate
can alter dramatically the behavior, showing that cultural drift,
as modelled by continuous random perturbations, has a relevant role
in the behavior of Axelrod's model.

Finally, notice that our argumentation based on relaxation times
of perturbations so far does not involve the value of $q$ and it
may explain the weak $q$--dependence of the system in the presence
of noise.

%%%%%%%%%%%%%%%%%%%%%%%%%%%%%%%%%%%%%%%%%%%%%%%%%%%%%%%%%%%%%%%%%%%%%
\section{Conclusions and Outlook}

We have shown that the ACM, a model of cultural transmission, in a
one-dimensional world can be understood as an optimization process in
which the global uniform state is the optimum state corresponding to
the global minimum of a Lyapunov potential. When the initial cultural
diversity is large enough the system freezes in an attractor of the
dynamics. The system can always escape from these attractors by any
small perturbation, since there are always nearby configurations with
the same or lower value of the potential.  Cultural drift gives rise to
such perturbations, and therefore it is an instrument to promote
cultural globalization giving to the system the necessary input to
proceed in the optimization dynamics. However, if cultural drift acts
at high enough rate it leads to a noisy polarized dynamical state.

As we have previously mentioned, we have only found a Lyapunov
potential for the one-dimensional version of the ACM. However, most of
our qualitative findings persist in simulations in a two-dimensional
regular network \cite{Klemm02a}. In the two-dimensional world there is
a transition between the uniform and multicultural states for a
critical value of $q$, but the attractors are not easily classified in
terms of the number of barriers, and the uniform state is not known to
be the global minimum of a potential. The dynamical stability of the
other attractors is also unknown. Still, simulations indicate that
perturbations acting on the polarized states take the system to the
uniform state. In the presence of cultural drift there is also a
transition from uniform states to a polarized multicultural state
controlled by the noise rate.

Our discussion has been here restricted to regular networks with
interactions between nearest neighbors. However, social networks are
known to be in many cases different from regular or random networks.
The question of the influence of network topologies reflecting social
cleavages was already posed by Axelrod \cite{Axelrod97a,Axelrod97b}.
Two types of networks very much studied recently are the small wold
networks \cite{Watts98}, representing an intermediate situation between
regular and random networks, and the scale free networks
\cite{Barabasi99a}, characterized by a power law tail in the
probability distribution for the number of bonds connecting of a site
in the network. Such power law indicates the presence of few sites with
a very large number of links. Simulations of the Axelrod model in these
networks \cite{Klemm02b} indicate that the small world connectivity
favors cultural globalization, in the sense that the value of $q_c$ for
the transition to a polarized multicultural state is larger than in the
regular network. A maximum value of $q_c$ is obtained for the random
network, but the scale free connectivity is still more efficient than
the random connectivity in promoting global culture, giving a larger
value of $q_c$. In fact this value depends on the system size, and in
the limit of a very large system size, the system reaches the uniform
multicultural state for any value of $q$. The interesting unsolved
question so far is to take into account that if the cultural evolution
of the individual is molded by the network of social interactions, the
network is also constructed by the individuals. Properly the network
can not be taken as given and fixed \cite{Lazer01}. Such co-evolution
of individual culture and social network could be modelled similarly to
studies of cooperation in which the social network emerges from the
results of the dynamics of cooperation \cite{Zimmermann01}.

%%%%%%%%%%%%%%%%%%%%%%%%%%%%%%%%%%%%%%%%%%%%%%%%%%%%%%%%%%%%%%%%%%%%%
% \section*{Acknowledgments}
% 
% We acknowledge financial support from MCyT (Spain) and FEDER (EU)
% through Projects CONOCE and BFM2002-04474-C02-01.

%%%%%%%%%%%%%%%%%%%%%%%%%%%%%%%%%%%%%%%%%%%%%%%%%%%%%%%%%%%%%%%%%%%
% \begin{thebibliography}{99}

\ignore{
\bibliographystyle{econometrica}
% \bibliography{/home/victor/tex/biblio/games}
\bibliography{games}

% \end{thebibliography}
}

\ifx\undefined\BySame
\newcommand{\BySame}{\leavevmode\rule[.5ex]{3em}{.5pt}\ }
\fi
\ifx\undefined\textsc
\newcommand{\textsc}[1]{{\sc #1}}
\newcommand{\emph}[1]{{\em #1\/}}
\let\tmpsmall\small
\renewcommand{\small}{\tmpsmall\sc}
\fi

\end{document}